\documentclass[10pt,conference]{IEEEtran}
\IEEEoverridecommandlockouts
\usepackage{cite}
\usepackage{amsmath,amssymb,amsfonts}
\usepackage{graphicx}
\usepackage{xcolor}
\usepackage[hyphens]{url}   
\usepackage{hyperref}

\usepackage{multirow}
\usepackage{booktabs}
\usepackage{subcaption}
\usepackage{enumitem}
\usepackage{pifont}
\usepackage{cleveref}   

\newcommand{\cmark}{\textcolor{green!60!black}{\ding{51}}}
\newcommand{\xmark}{\textcolor{red!70!black}{\ding{55}}}

\newcommand{\az}[1]{}
\newcommand{\gw}[1]{}
\newcommand{\pgf}[1]{}

\title{When Does Disaggregation Pay? Simulating Prefill--Decode--Attention--FFN Specialization for Agentic LLM Inference}

\hypersetup{
  pdftitle={When Does Disaggregation Pay? Simulating Prefill-Decode-Attention-FFN Specialization for Agentic LLM Inference},
  pdfauthor={Przemyslaw Forys, Haoran Wu, Can Xiao, Jiayi Nie, Tony Liu, Rika Antonova, Timothy Jones, Robert Mullins, Wayne Luk, Aaron Zhao, George A. Constantinides},
  pdfsubject={Computer architecture; machine learning systems},
  pdfkeywords={LLM inference, disaggregated serving, agentic inference, NPU, heterogeneous systems, simulation},
  colorlinks=false,
  pdfborder={0 0 0}
}

\author{%
  \IEEEauthorblockN{%
    Przemyslaw Forys\IEEEauthorrefmark{1},
    Haoran Wu\IEEEauthorrefmark{2},
    Can Xiao\IEEEauthorrefmark{1},
    Jiayi Nie\IEEEauthorrefmark{2},
    Tony Liu\IEEEauthorrefmark{1},
    Rika Antonova\IEEEauthorrefmark{2},\\
    Timothy Jones\IEEEauthorrefmark{2},
    Robert Mullins\IEEEauthorrefmark{2},
    Wayne Luk\IEEEauthorrefmark{1},
    Aaron Zhao\IEEEauthorrefmark{1},
    George A. Constantinides\IEEEauthorrefmark{1}%
  }
  \IEEEauthorblockA{%
    \IEEEauthorrefmark{1}Imperial College London, UK \qquad
    \IEEEauthorrefmark{2}University of Cambridge, UK\\[2pt]
    \footnotesize
  }%
}

\begin{document}
\maketitle

\thispagestyle{plain}
\pagestyle{plain}


\begin{abstract}

Agentic inference now dominates the LLM inference landscape, requiring LLMs to actively engage in multi-turn interactions with tool-calling capabilities.
This introduces a more complex workload for the underlying inference system: serving stages such as prefill and decode exhibit substantially different behaviors and demand distinct compute and memory-bandwidth capabilities.
As a result, a single homogeneous GPU system now struggles to support agentic inference, motivating an industry shift toward heterogeneous systems with disaggregated serving capabilities, such as the emerging Vera-Rubin platform with GPUs and Groq LPUs.
However, the question of what the optimal hardware should look like for each component in a heterogeneous system remains underexplored.
To this end, we propose a novel simulation framework for disaggregated serving, termed \textbf{HeteroPanacea}, that enables system-level simulation across three dimensions: 1) disaggregated quantization, 2) automated intra- and inter-device parallelization scheduling, and 3) PDAF (prefill-decode-attention-FFN) NPU architectural heterogeneity.
By combining these three axes, we provide a cross-stack simulation framework for future heterogeneous agentic serving systems.
We confirm the benefit of Prefill Decode disaggregation, simulating increased serving throughput by up to 75\% compared to traditional serving with current GPUs and demonstrate 4 way Prefill Decode Attention FFN disaggregation is the most consistent for increasing throughput across different models, assuming custom NPUs. We also investigate the relationship between model architecture and gain from disaggregation by running a set of ablation studies.

\end{abstract}

\section{Introduction}

\begin{figure}[t]
    \centering

    \begin{subfigure}[t]{0.33\linewidth}
        \centering
        \includegraphics[width=\linewidth]{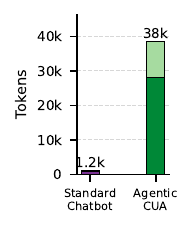}
        \caption{Agentic workloads have larger token usage.}
        \label{fig:token_usage}
    \end{subfigure}
    \hfill
    \begin{subfigure}[t]{0.64\linewidth}
        \centering
        \includegraphics[width=\linewidth]{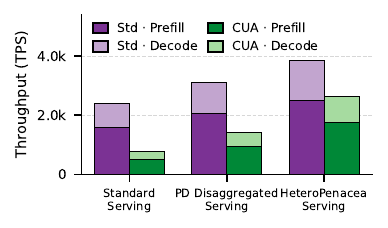}
        \caption{System performance comparison of different serving systems,
        split into prefill and decode throughput.}
        \label{fig:perf_comparison}
    \end{subfigure}

    \vspace{0.75em}

    \begin{subfigure}[t]{\linewidth}
        \centering
        \includegraphics[width=\linewidth]{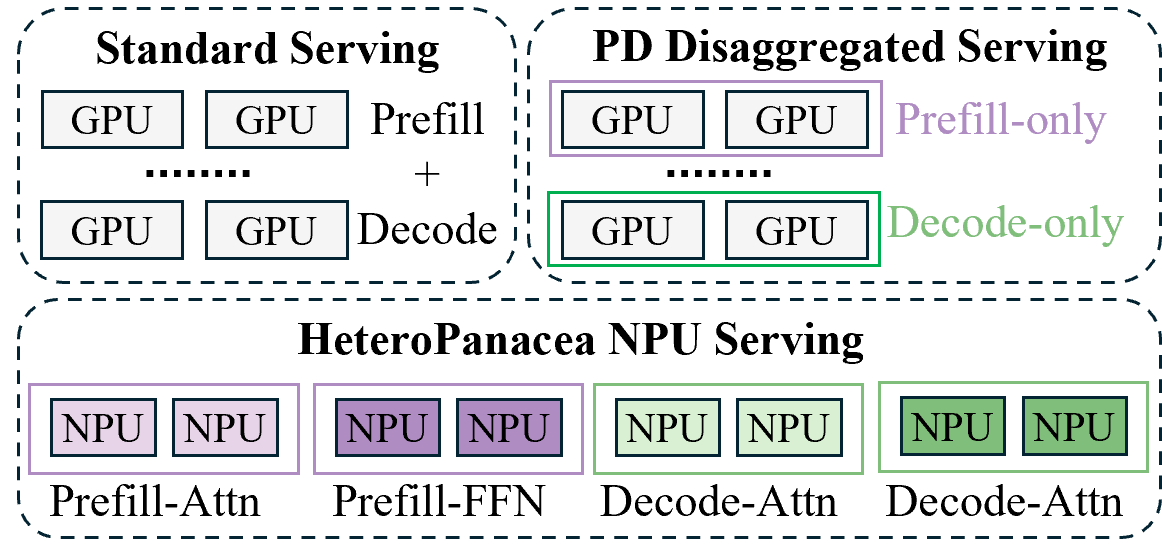}
        \caption{Feature comparison of standard serving, homogeneous PD-disaggregated serving, and HeteroPanacea NPU serving with stage-optimized configurations, indicated by different colors.}
        \label{fig:third_motivation}
    \end{subfigure}

    \caption{Motivation of HeteroPanacea, a heterogeneous NPU serving system for agentic LLM workloads.}
    \label{fig:combined_motivation}
\end{figure}

Large language models (LLMs) are increasingly moving beyond single-turn chatbot interactions toward agentic applications, where models reason, act, and interact with external environments over multiple turns. Representative workloads include computer-use agents (CUAs)~\cite{OSWorld}, autonomous coding agents~\cite{kernelcraft,rando2025longcodebench}, and web-use agents~\cite{OSWorld,he2024webvoyager,webagent}.
During these interactions, screenshots, web content, code context, intermediate reasoning, tool calls, and user feedback are repeatedly accumulated into the prompt, placing substantially higher memory and compute demands on the serving system than traditional chatbot inference~\cite{gsm8k}. Context length consequently grows rapidly during inference: on the OSWorld~\cite{OSWorld} benchmark it averages $38$K tokens and can reach $100$K, more than an order of magnitude beyond a standard chatbot session, as shown in \Cref{fig:token_usage}.

As context length grows, the architectural mismatch between the distinct stages of the inference pipeline becomes increasingly pronounced on current AI inference devices, and a single homogeneous device can no longer efficiently sustain the end-to-end inference process. Prefill-Decode (PD) disaggregation has consequently emerged as a popular architectural paradigm in modern AI accelerator design~\cite{distserve,li2026prefillsequalppddisaggregation}. As illustrated in \Cref{fig:perf_comparison}, where all systems are simulated under the same power budget and request rate, PD disaggregation delivers up to a $1.82\times$ throughput improvement over a unified execution model on agentic workloads, against $1.29\times$ on a standard chatbot workload. Building on this paradigm, recent systems push disaggregation further by tailoring the hardware to each stage. NVIDIA's upcoming Vera Rubin platform, for example, pairs Rubin GPUs -- which handle the prefill stage and attention computation during decode -- with Groq LPUs dedicated to FFN computation during decode~\cite{vera_rubin}.

\begin{table*}[t]
\centering
\caption{Feature comparison of representative LLM inference simulation frameworks. HeteroPanacea uniquely combines heterogeneous hardware modeling, PD and AF disaggregation, disaggregated quantization, multiple parallelization strategies, and design-space search.}
\label{tab:feature_comparison}
\setlength{\tabcolsep}{6pt}
\renewcommand{\arraystretch}{1.15}
\begin{tabular}{lcccccc}
\toprule
\textbf{Framework}
& \textbf{Heterogeneity}
& \textbf{PD Disagg.}
& \textbf{AF Disagg.}
& \textbf{Disagg. Quant.}
& \textbf{Parallelism}
& \textbf{Search}\\
\midrule
LLMCompass~\cite{llmcompass}
& \xmark & \xmark & \xmark & \xmark & DP & \cmark\\

MemExplorer~\cite{memexplorer}
& \cmark & \cmark & \cmark & \xmark & TP/DP/PP/EP & \xmark \\

DistServe~\cite{distserve}
& \xmark & \cmark & \xmark & \xmark & TP/DP/PP & \cmark \\
LLMServingSim2.0~\cite{cho2026llmservingsim20unifiedsimulator}
& \cmark & \cmark & \cmark & \xmark & TP/DP/PP/EP & \xmark \\

\textbf{HeteroPanacea}
& \cmark & \cmark & \cmark & \cmark & TP/DP/PP/EP & \cmark \\
\bottomrule
\end{tabular}
\end{table*}

However, despite the rapid industry shift toward heterogeneous, disaggregated systems, the question of what the optimal hardware should look like for each disaggregation stage remains largely underexplored, precisely because the \textit{NPU architecture best suited to one disaggregation stage can look nothing like the one best suited to the next}.
Two questions follow: to what extent does PD disaggregation remain beneficial under different workload profiles, and how do software-level optimizations such as quantization interact with, and shift, the optimal hardware configuration for each stage?
Answering them -- that is, systematically exploring this hardware design space -- is non-trivial.
It requires jointly reasoning across multiple, tightly coupled dimensions: the quantization chosen for each disaggregated stage; the parallelization strategy (e.g., tensor, pipeline, or expert parallelism) applied independently to prefill versus decode; the underlying hardware architecture selected for prefill versus decode NPUs; and even finer-grained heterogeneity within a single stage, such as assigning different hardware or precision to Attention versus FFN layers (also known as AF disaggregation). Existing simulation tools, summarized in \Cref{tab:feature_comparison}, treat these dimensions in isolation, if at all.

To address these issues, we propose \textbf{HeteroPanacea}, a simulator for disaggregated, heterogeneous inference systems serving agentic workloads. HeteroPanacea enables system-level design-space exploration along five axes: 
(i) data, tensor, and pipeline parallelism; 
(ii) Prefill-Decode disaggregation; 
(iii) Attention-FFN disaggregation; 
(iv) fine-grained heterogeneous NPU configurations (e.g., matrix-engine size, SRAM capacity); 
and (v) mixed-precision quantization. 
Beyond throughput and latency, HeteroPanacea also models the accuracy impact of stage-wise quantization: attention and linear layers in the prefill and decode stages can each be assigned independent MXINT configurations, enabling fine-grained accuracy-efficiency trade-off analysis. 
We use HeteroPanacea to study the design principles underlying next-generation AI infrastructure for agentic workloads, and will open-source the framework in full -- simulator, configurations, and evaluation scripts -- upon acceptance.
The main contributions are as follows:

\begin{itemize}[leftmargin=1em]
    \item We propose \textbf{HeteroPanacea}, a heterogeneous system-level simulation framework for agentic LLM inference. HeteroPanacea supports configurable exploration of parallelization strategies, heterogeneous NPU architectures, mixed-precision configurations, and disaggregated serving designs.

    \item We demonstrate the flexibility of the simulation framework by systematically evaluating disaggregated serving across diverse model architectures and workload profiles, and by benchmarking conventional Prefill-Decode and Attention-FFN disaggregation against a novel four-stage \textbf{PDAF} architecture (Prefill/Decode $\times$ Attention/FFN) that we introduce, which jointly disaggregates both dimensions. In HeteroPanacea, the NPU design at each disaggregation stage can be independently customized: Prefill-Attention, Prefill-FFN, Decode-Attention, and Decode-FFN can each be assigned a distinct NPU hardware architecture, as illustrated in \Cref{fig:third_motivation}.

    \item We characterize the limitations of homogeneous hardware for disaggregated serving and investigate fully heterogeneous systems in which each inference stage is mapped to a stage-specialized NPU configuration. In our experiments, PD and PDAF disaggregation both benefit from specialized hardware for each stage, achieving up to a $2.06\times$ throughput gain on agentic workloads compared to traditional serving. Crucially, this gain is conditional: disaggregation only clears parity once the workload is prefill-heavy, and the four-way PDAF split only outperforms plain PD when the hardware design space is rich enough to give attention and FFN genuinely different devices.
    
\end{itemize}

\section{Background and Related Work}
\subsection{Disaggregated Inference Serving}

Modern LLMs, built on the Transformer architecture~\cite{vaswani2017attention}, generate text autoregressively: given an input prompt, the model first processes all input tokens in a parallelizable prefill phase, then generates output tokens one at a time in an iterative decode phase. These two phases exhibit fundamentally different computational characteristics -- prefill is compute-bound and amenable to high GPU utilization, while decode is memory-bandwidth-bound due to the sequential nature of token generation and the growing key-value (KV) cache~\cite{distserve}.

As serving systems scale to handle thousands of concurrent requests, the heterogeneity between these computational phases creates resource contention and latency inefficiencies. A growing body of work has therefore explored disaggregated LLM serving -- architectures that decouple different phases or components of inference onto separate hardware pools -- in order to independently optimize throughput, latency, and resource utilization for each stage~\cite{zhong2024distserve, patel2023splitwise}.

\emph{Prefill--decode (PD) disaggregation} places the prefill and decode phases on separate groups of serving devices.
In conventional co-located serving, prefill requests interfere with decode iterations, increasing both TTFT and TPOT.
Isolating and pipelining the two phases lets the prefill pool be optimised for TTFT and compute throughput while the decode pool is optimised for TPOT and memory efficiency.
Representative systems such as DistServe~\cite{zhong2024distserve} and Splitwise~\cite{patel2023splitwise} show that this separation improves service-level objective (SLO) attainment and achieves better goodput.
Two system-level problems arise from this separation.
First, the prompt's KV must be transferred from the prefill instance to the decode instance over the network, so the handoff cost is set by the topology and link bandwidth rather than by the model.
Second, the two phases do not proceed at the same rate, so the pools must be provisioned and load-balanced against each other --- prefill running ahead backs KV up behind decode, prefill lagging leaves the decode devices idle --- and the ratio that balances them follows a prompt-to-output length mix that shifts at runtime.
Simulating PD is therefore no longer an intra-replica scheduling problem but an inter-replica routing and bandwidth one.

\emph{Attention--FFN (AF) disaggregation} goes one level finer and splits the decode phase itself.
It places the attention and FFN modules on separate GPU groups.
Batching moves the two in different directions.
Attention reads a \emph{distinct} KV cache per request, so a larger batch adds memory traffic without adding reuse and the operator stays memory-bandwidth-bound.
The FFN instead applies the \emph{same} weights to every token it serves, so a larger batch enables more reuse.
AF introduces two costs of its own.
First, hidden states cross between the groups at every layer, which introduces smaller and more frequent network traffic.
Second, the groups run as a pipeline over several microbatches, and their per-stage latencies must be matched almost exactly, since any imbalance stalls the pipeline and leaves one side idle~\cite{step3}.
Simulating AF therefore requires accurate modelling of per-layer inter-group traffic, together with the pipeline balance it depends on.

In this paper, we focus on a four-way PDAF disaggregation, a finer-grained scheme that decomposes inference into four distinct stages, prefill-attention, prefill-FFN, decode-attention, and decode-FFN, which we refer to as \textit{disaggregation stages} throughout the paper. It is worth noting that PDAF subsumes the two coarser-grained schemes as special cases: merging attention and FFN within each of the prefill and decode stages recovers conventional PD disaggregation, while merging prefill and decode within each of the attention and FFN stages recovers conventional AF disaggregation.

\subsection{Heterogeneous Accelerators and Memory Hierarchies}

Different inference stages have distinct compute, bandwidth, and capacity requirements, motivating heterogeneous accelerator and memory designs. Prior simulation frameworks such as LLMServingSim 2.0~\cite{cho2026llmservingsim20unifiedsimulator} and MemExplorer~\cite{memexplorer} explore heterogeneous serving systems. However, LLMServingSim 2.0 focuses on fixed hardware and relies on profiling, while MemExplorer does not model distributed parallelism. HeteroPanacea instead enables exploration of custom NPUs, parallelization strategies, disaggregation, and mixed-precision quantization.

Several accelerator architectures have been proposed for efficient neural-network and LLM inference, including PLENA~\cite{plena}, FlightLLM~\cite{flightllm}, MicroscalingQ~\cite{Microscopiq}, and the Coral NPU~\cite{google2025_coral_npu}. We adopt PLENA as the baseline compute architecture because it provides a highly parameterizable and representative NPU substrate rather than a fixed accelerator instance. 

\section{Motivation}

\label{sec:motivation}
Prefill and decode place fundamentally different demands on hardware: prefill is compute-bound at long sequence lengths, while decode is bound by the bandwidth needed to read KV cache and model weights for a single new token per user request. This mismatch is well understood and has already driven industry adoption of prefill/decode (PD) disaggregation, where the two phases run in separate hardware pools that are sized independently \cite{zhong2024distserve, patel2023splitwise, qin2025mooncake}. PD is, however, only the coarsest cut through the inference pipeline: it treats each phase as a monolithic unit. In fact, prefill and decode each contain two compute-heavy sub-layers, attention and feed-forward (FFN), whose resource profiles diverge from each other just as sharply as prefill diverges from decode, though the character of that divergence differs by phase.

In decode, attention's cost is dominated by reading a KV cache that grows with
context length, making it a bandwidth-bound, capacity-hungry sub-stage. Because
MoE sparsity lives entirely in the FFN, this is independent of sparsity; the KV
footprint does depend heavily on the attention variant (MHA, GQA, or
MLA~\cite{deepseekv3}), but that axis is orthogonal to sparsity. Decode-FFN's
cost, by contrast, is dominated by streaming weight matrices, and in
mixture-of-experts (MoE) models only a small fraction of experts activate per
token, so reaching high FFN utilization requires aggregating a batch large
enough that each active expert sees enough tokens, and the sparser the routing,
the larger that batch must be~\cite{step3}. At any fixed serving batch, then,
FFN's effective arithmetic intensity falls as sparsity increases, pushing it
toward a different point on the compute/bandwidth roofline than attention. In
prefill both sub-layers are compute-bound rather than bandwidth-bound, but their
intensities still diverge: attention compute grows quadratically with sequence
length (the $O(n^{2})$ $QK^{\top}$ and $\mathrm{softmax}(\cdot)V$ terms), while
FFN is a dense GEMM whose cost scales with token count and, for MoE, with
routing. The attention/FFN mismatch is thus real within each phase, not only
across the prefill/decode boundary.

Under PD, the ``decode'' node still forces decode-attention (DA) and decode-FFN (DF), two sub-stages with different ideal compute-to-bandwidth ratios, onto the same physical device. Whatever hardware is chosen sizes correctly for one and wastes capacity on the other. This residual, within-phase mismatch is exactly what motivates us to go one step further than PD to full four-way disaggregation (PDAF): separating prefill-attention, prefill-FFN, decode-attention, and decode-FFN onto independently specialized hardware pools. Disaggregating attention from FFN has been shown to enable independent scaling and heterogeneous deployment of the two sub-layers~\cite{step3}; PDAF extends that split across the prefill/decode boundary as well.

This residual mismatch is not static: it is widening. Growing context lengths, driven by long-document processing, long-chain-of-thought reasoning, and especially the rapid rise of agentic workloads (repeated tool calls and retrievals that re-prefill an ever-growing context on nearly every turn), inflate KV-cache volume and therefore decode-attention's bandwidth demand, while decode-FFN's cost is governed instead by expert count and routing sparsity. As these two pressures increasingly diverge, the case for separating DA from DF only strengthens, and the same argument applies on the prefill side as prompts lengthen -- there through attention's quadratic compute rather than KV bandwidth. PD's coarse split cannot capture this; only stage-level disaggregation can.

Realizing PDAF, however, multiplies the design space: each of the four sub-stages \textit{can now be given its own hardware} (compute throughput, memory capacity and bandwidth, interconnect), parallelism strategy, and replica count, under a shared power or cost budget, and the benefit of this extra granularity is not guaranteed to be worth its complexity: it depends on model architecture and, as we show, on the workload's prefill/output ratio. Evaluating this joint space empirically, across enough hardware and workload points to know when PDAF's extra specialization pays for itself over plain PD, is infeasible on real clusters. This motivates a fast, analytically grounded simulator that can sweep this space cheaply and identify the conditions under which the fourth-way split, attention from FFN rather than just prefill from decode, is actually worth deploying.

\section{Inference Simulation}
This section describes the models used to simulate inference, and the experiments run to validate them.

\subsection{Validation Platform}
\label{sec:validation-platform}

Measurements in this section were collected on a single
server: 8$\times$ NVIDIA B200, Intel Xeon 6960P, CUDA 12.8, PyTorch 2.10.0+cu128,
NCCL 2.27.5, Python 3.12.13. Compute kernels use cuBLAS/cuBLASLt via
\texttt{torch.matmul} (BF16) and \texttt{torch.\_scaled\_mm} (FP8~E4M3);
collectives use NCCL. Timings are CUDA-event based with warm-up and an
adaptive iteration count sized to a fixed wall-clock budget.

\subsection{Compute Device (NPU) Model}
\label{sec:sim-npu-model}

Rather than a cycle-accurate microarchitectural model, each accelerator in
the design space is characterized by a small set of
peak specifications and a roofline execution model built on top of them.
This keeps the design space tractable 
while still capturing the two resources that determine inference latency:
compute throughput and memory bandwidth.

\begin{table}[!t]
\centering
\caption{Device design space explored in the hardware sweep. Each memory
  technology contributes 2--5 representative (capacity, bandwidth) operating
  points.}
\label{tab:design-space}
\small
\setlength{\tabcolsep}{4pt}
\begin{tabular}{@{}lrr@{}}
\toprule
\multicolumn{3}{@{}l@{}}{%
  \textbf{TFLOPS:}\enspace
  25\quad 100\quad 250\quad 2{,}500\quad 10{,}000\quad 20{,}000} \\
\midrule
\textbf{Memory Type} & \textbf{Capacity (GB)} & \textbf{Bandwidth (GB/s)} \\
\midrule
\multirow{2}{*}{SRAM}  &   0.1 &  1{,}000 \\
                       &   0.1 & 10{,}000 \\
\cmidrule{1-3}
\multirow{5}{*}{HBM}   &  16   &  1{,}000 \\
                       &  40   &  2{,}000 \\
                       &  80   &  3{,}500 \\
                       & 140   &  4{,}800 \\
                       & 190   &  8{,}000 \\
\cmidrule{1-3}
\multirow{2}{*}{DDR}   &   4   &     20   \\
                       &  16   &     60   \\
\cmidrule{1-3}
\multirow{2}{*}{LPDDR} &   8   &    100   \\
                       &  32   &    200   \\
\cmidrule{1-3}
\multirow{2}{*}{GDDR}  &   4   &    300   \\
                       &  16   &    600   \\
\bottomrule
\end{tabular}
\end{table}

\paragraph{Device parameterization.}
A device is described by a peak compute rate $F$ (TFLOPS, precision-agnostic
peak), an \emph{exclusive} memory technology class drawn from seven
candidates, and that class's paired per-device capacity $C$ (GB) and
bandwidth $B$ (GB/s). The memory subsystem is
parameterized by representative $(C, B)$ points drawn from each
technology's empirical design space rather than a free-form continuous
sweep. \Cref{tab:design-space} lists the full device parameter set; the
parallelism degrees each pool may be configured with are described in
\Cref{sec:sim-parallelism-model}.

\begin{figure*}
    \centering
    \includegraphics[width=1\linewidth]{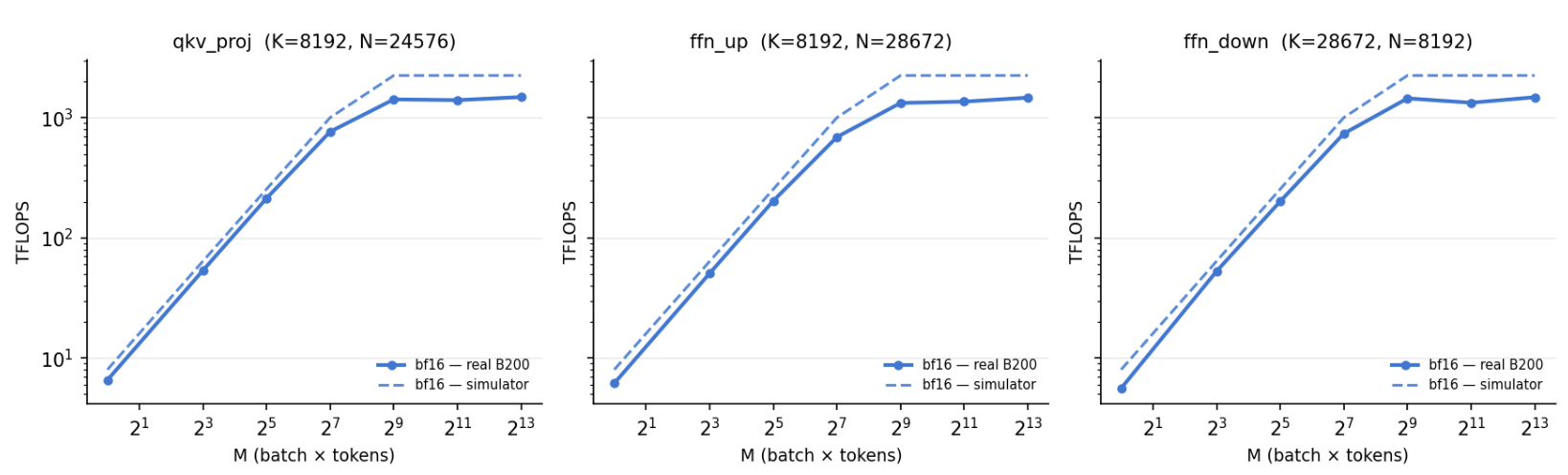}
    \caption{Comparison of achieved TFLOPS of the roofline model compared with experimental data}
    \label{fig:cv}
\end{figure*}

\paragraph{Roofline execution time.}
Every stage of inference (prefill-attention, prefill-FFN, decode-attention,
decode-FFN) is decomposed into a FLOP count and a byte count from the
model's architecture (accounting for GQA/MHA vs.\ multi-head latent
attention (MLA) KV-cache layout, and dense vs.\ mixture-of-experts (MoE)
FFN routing), and executed under the standard roofline bound:
\begin{equation}
  \label{eq:roofline}
  t = \max\!\left(\frac{\Phi}{F_{\mathrm{eff}}},\; \frac{\beta}{B_{\mathrm{eff}}}\right),
\end{equation}
where $\Phi$ and $\beta$ are the operation's FLOP and byte counts and
$F_{\mathrm{eff}}, B_{\mathrm{eff}}$ are the \emph{effective} compute rate
and bandwidth after accounting for tensor- and pipeline-parallel scaling
(\Cref{sec:sim-parallelism-model}). No queuing, kernel-launch, or
occupancy effects are modeled below this per-stage granularity. The achieved
TFLOPS predicted by the roofline model are compared against measured data in
\Cref{fig:cv}.

\paragraph{Power.}
For GPU-based systems we use the power reported in the datasheet. NPU device power is modeled using data from existing GPU specifications, as described below.

Device power is the sum of two independently modeled terms,
$P = P_{\mathrm{compute}} + P_{\mathrm{mem}}$, evaluated at each device's
peak specification. $P_{\mathrm{compute}}$ follows a
sub-linear power law, $P_{\mathrm{compute}} = P_{\mathrm{ref}}\,(F /
F_{\mathrm{ref}})^{\alpha}$, where the reference point
$(P_{\mathrm{ref}}, F_{\mathrm{ref}})$ is anchored to the H100 datasheet and
the exponent $\alpha$ is fit to published TDP minus memory-subsystem power
across three real accelerators spanning three device generations. The
sub-linear exponent reflects diminishing marginal power cost per FLOP as
process node and architecture improve.

\begin{figure}
    \centering
    \includegraphics[width=1\linewidth]{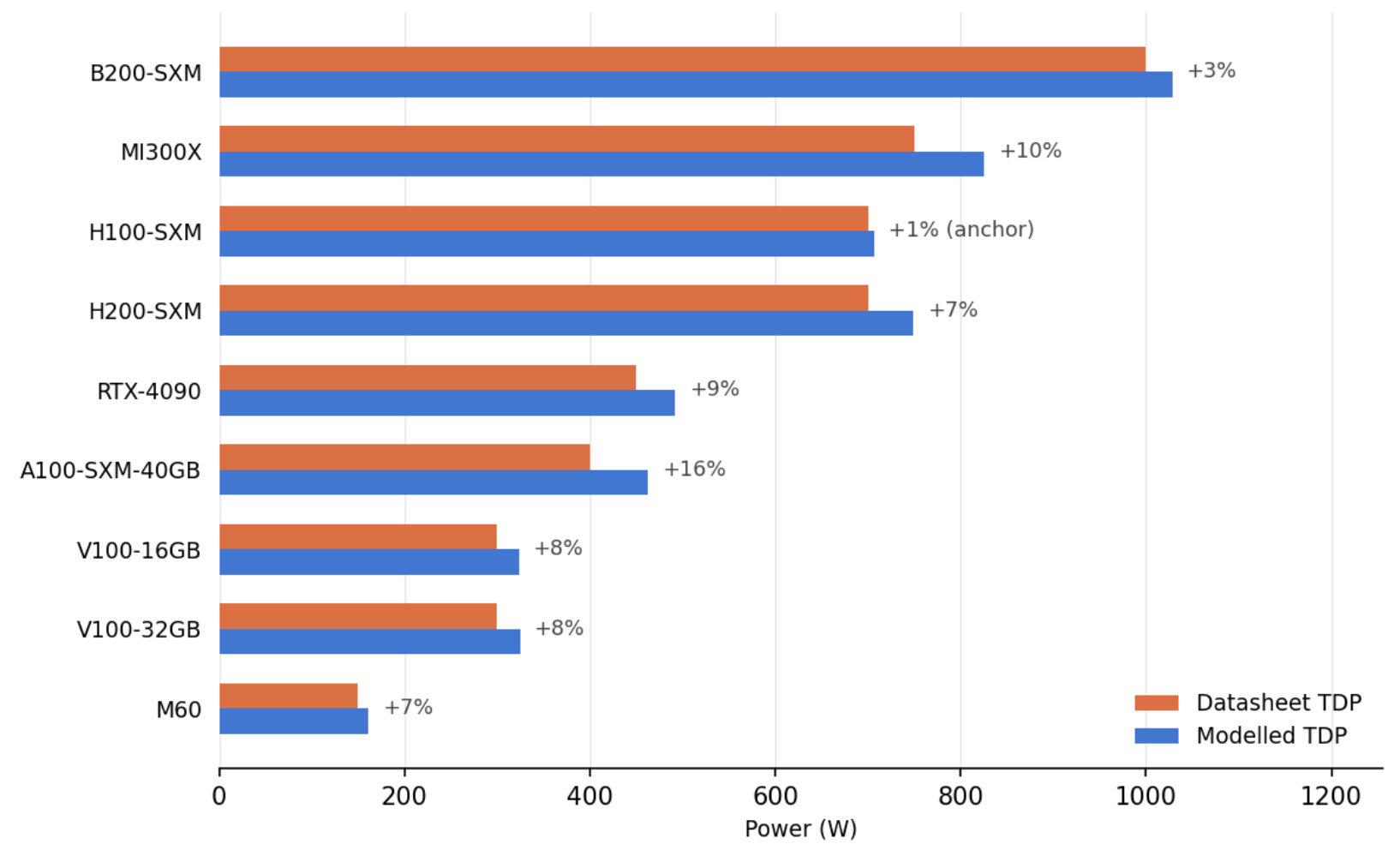}
    \caption{Comparison of simulated power of GPUs with datasheet TDP}
    \label{fig:pv}
\end{figure}

$P_{\mathrm{mem}}$ is a
physics-based dynamic-plus-leakage estimate per memory technology, adapted from MemExplorer~\cite{memexplorer}: on-chip
technologies (SRAM, 3D-SRAM) use an on-chip power estimator; off-chip
technologies (HBM, HBF, DDR, LPDDR, GDDR) use per-bit read/write energy
 scaled by bandwidth for the dynamic term, plus an idle-mode power
density scaled by capacity for the leakage term. We compare simulated power with actual TDP of GPUs in \Cref{fig:pv}.

\subsection{Interconnect Model: Device-to-Device (D2D) and Node-to-Node (N2N)}
\label{sec:sim-interconnect-model}

Communication is modeled with the same roofline philosophy as compute: a
byte count derived analytically from the operation being performed, divided
by a bandwidth appropriate to \emph{where} that transfer physically
occurs. Two distinct bandwidth domains are exposed per device:

\begin{itemize}
  \item \textbf{D2D} intra-node,
        device-to-device bandwidth. This is the bandwidth used for tensor- and
        expert-parallel collectives and for pipeline-parallel activation
        hand-off between stages co-located on the same node.
  \item \textbf{N2N} inter-node bandwidth, used specifically for the KV-cache
        and activation transfers that cross a \emph{disaggregation} boundary
        (\Cref{sec:sim-parallelism-model}) between physically
        separate stage pools.
\end{itemize}

\begin{table}[t]
\centering
\small
\caption{%
  Measured versus simulated latency for intra-op (tensor) and inter-op
  (pipeline) parallelism, $L=8$ layers, $d_{\mathrm{model}}=8192$,
  $M=2048$ tokens. Simulated values use each mode's own calibrated bandwidth:
  a measured \texttt{all\_reduce} rate for TP and a point-to-point rate
  for PP.}
\label{tab:parallelism_scaling}
\begin{tabular}{llrrr}
\toprule
\textbf{Mode} & \textbf{\#GPUs} & \textbf{Measured} & \textbf{Simulated} & \textbf{Sim/Real} \\
 &  & (ms) & (ms) & \\
\midrule
\multirow{4}{*}{Intra-op (TP)} & 1 & 1.23 & 0.98 & 0.79$\times$ \\
 & 2 & 34.47 & 34.33 & 1.00$\times$ \\
 & 4 & 32.62 & 34.44 & 1.06$\times$ \\
 & 8 & 38.16 & 39.05 & 1.02$\times$ \\
\midrule
\multirow{4}{*}{Inter-op (PP)} & 1 & 1.45 & 0.98 & 0.67$\times$ \\
 & 2 & 2.42 & 1.85 & 0.77$\times$ \\
 & 4 & 4.32 & 3.61 & 0.84$\times$ \\
 & 8 & 7.90 & 7.12 & 0.90$\times$ \\
\bottomrule
\end{tabular}
\end{table}

\begin{table}[t]
\centering
\small
\caption{%
  Measured versus simulated latency for two communication primitives}
\label{tab:comm_validation}
\begin{tabular}{llrrr}
\toprule
\textbf{Primitive} & \textbf{Tokens} & \textbf{Measured} & \textbf{Simulated} & \textbf{Sim/Real} \\
 &  & ($\mu$s) & ($\mu$s) & \\
\midrule

\multirow{3}{*}{PP point-to-point} & 128 & 67 & 55 & 0.82$\times$ \\
 & 1024 & 449 & 437 & 0.97$\times$ \\
 & 8192 & 3,507 & 3,495 & 1.00$\times$ \\
\midrule
\multirow{3}{*}{EP all-to-all} & 128 & 63 & 55 & 0.86$\times$ \\
 & 1024 & 383 & 437 & 1.14$\times$ \\
 & 8192 & 2,891 & 3,495 & 1.21$\times$ \\
\bottomrule
\end{tabular}
\end{table}

\begin{figure*}
    \centering
    \includegraphics[width=1\linewidth]{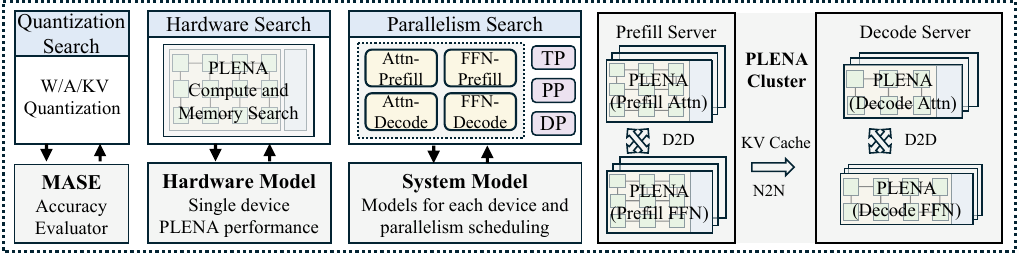}
    \caption{Overview of HeteroPanacea. TP - Tensor Parallelism, PP - Pipeline Parallelism, DP - Data parallelism, D2D - Device to device, N2N - Network to network}
    \label{fig:overview}
\end{figure*}

\subsection{Parallelism Model}
\label{sec:sim-parallelism-model}

Each stage pool (\Cref{sec:sim-npu-model}) is configured with four
independent parallelism degrees: tensor (TP, $t$), pipeline (PP, $p$),
data (DP, $d$), and expert (EP, $e$) subject to $t \cdot p \cdot d \le$
the number of physical devices assigned to that pool. Parallelism is
orthogonal to \emph{stage disaggregation}: each
pool independently chooses its own $(t, p, d, e)$ and device allocation.

\paragraph{Tensor parallelism (TP).}
Each of the $t$ devices in a TP group holds $1/t$ of every weight matrix
(row- or column-sharded). Aggregate compute and bandwidth for the group
therefore scale \emph{linearly} with $t$, at the cost of a
ring all-reduce after every attention output projection and every FFN
down-projection (\Cref{sec:sim-interconnect-model}).

\paragraph{Pipeline parallelism (PP).}
The model's layers are split into $p$ sequential stages, each holding
$L/p$ layers on separate devices. 
Every inter-stage boundary additionally requires one point-to-point
activation transfer (\Cref{sec:sim-interconnect-model}). Unlike TP,
PP does \emph{not} reduce per-request compute; its throughput
benefit comes entirely from overlapping micro-batches across stages, not
from per-request speedup.

\paragraph{Expert parallelism (EP).}
For MoE FFN layers, the $E$ experts are sharded $E/e$ per device. A token
activated for $k$ experts (top-$k$ routing) must be dispatched to whichever
device(s) own its routed experts and its output combined afterward,
contributing two all-to-all collectives per MoE layer whose volume scales
with the fraction of experts \emph{not} co-located with the token,
$(e-1)/e$.

\paragraph{Data parallelism (DP).}
DP is handled outside the roofline/communication model entirely: $d$
independent replicas, each a full copy of that stage's model and hardware
allocation, are instantiated as separate workers with disjoint request
queues at the scheduler level. DP therefore has no communication term and no
effect on per-request roofline time; it affects only aggregate throughput
(via the number of independent servers) and memory footprint (weights
are replicated $d$ times).

We validate the latency of communication primitives in
\Cref{tab:comm_validation}, and of the parallelism modes built on them in
\Cref{tab:parallelism_scaling}.

\subsection{Simulator}

We combine the models above into an end-to-end inference simulator.

\textbf{Disaggregation modes.} The simulator supports four deployment topologies of increasing stage separation: no disaggregation (prefill and decode share hardware), prefill/decode disaggregation, attention/FFN disaggregation, and a fully disaggregated mode separating all four sub-stages (prefill-attention, prefill-FFN, decode-attention, decode-FFN) onto independent hardware pools. Each mode defines a fixed routing graph between worker pools, with explicit KV-cache and activation transfers modeled wherever stages are physically separated.

\textbf{Scheduling.} The simulator advances through discrete events (arrivals, batch formation, stage completions, transfers, decode steps). Memory-aware continuous-batching schedulers govern prefill (greedy accumulation with timeout-based flushing) and decode (FIFO admission bounded by available KV-cache capacity).

\section{HeteroPanacea Search}

\subsection{Quantization search}
\label{sec:quant-search}

As a first method of optimizing the serving system, we explore the feasibility of reducing the precision of certain compute stages in order to reduce the compute and bandwidth required. A layer is only quantized if doing so results in no accuracy loss on the benchmark. Unlike the hardware search, precision cannot be ranked
analytically, since accuracy loss is only observable by running the
quantized model; this search is therefore empirical rather than
roofline-driven.

\textbf{Search space.} We assign an independent element bit-width
$q_s \in \mathcal{Q}$ to each of the four stages
$s \in \{\mathrm{PA}, \mathrm{PF}, \mathrm{DA}, \mathrm{DF}\}$, giving a
precision assignment $\mathbf{q} = (q_{\mathrm{PA}}, q_{\mathrm{PF}},
q_{\mathrm{DA}}, q_{\mathrm{DF}}) \in \mathcal{Q}^4$
(e.g.\ $\mathcal{Q} = \{4, 8, 16\}$). Both weights and activations are
quantized to the MXint format: a microscaling integer format sharing one
exponent across a block of $B$ elements, with $w_s$ set independently per tensor role (weights,
activations, KV cache) and, within attention, per sub-operation
($QK^\top$, $AV$) \cite{rouhani2023microscalingdataformatsdeep}. A single stage's precision is therefore itself a small
vector of block/width settings rather than one scalar, which we summarize
as $w_s$ for the search.

\textbf{Enforcing the assignment.} Applying $\mathbf{q}$ requires
distinguishing quantization along two axes: by layer type (attention vs.\
FFN) and by inference phase (prefill vs.\ decode). The former is handled
natively by the MASE~\cite{mase} quantization pass, which applies a
distinct config per matched layer pattern. The latter is not: MASE has no
notion of prefill vs.\ decode, so we attach a \emph{PhaseAutoSwitch} hook
to each quantized layer that inspects the sequence length $L$ of the
tensor passing through it ($L>1 \Rightarrow$ prefill config,
$L=1 \Rightarrow$ decode config) and switches to the corresponding
per-phase width at runtime.

\textbf{Search procedure.} For each candidate $\mathbf{q} \in
\mathcal{Q}^4$ we serve the quantized model and evaluate task accuracy
$A(\mathbf{q})$ against a held-out benchmark suite. We evaluate uniform fp16 and fp8 assignments as baselines, and compare them against settings in which one stage is reduced to 4 bits with the remaining three fixed at 8.

\subsection{Hardware and parallelism search}
\label{sec:workload-sweep}

Given a workload specification, the search finds the optimal hardware and parallelism setup.

It proceeds in two phases. \textbf{Phase 1} ranks hardware
candidates per stage analytically, without simulation, using the roofline
throughput model to compute a capacity score
\begin{equation}
    \sigma(\mathrm{hw}) = d \cdot \frac{\mathrm{tput}(B)}{\Lambda_s},
\end{equation}
the provisioned throughput of $d$ DP replicas divided by the stage's demand
rate $\Lambda_s$ ($\lambda$ for prefill stages, $\lambda O$ for decode
stages). Candidates exceeding $P_{\mathrm{budget}}$ or unable to hold model
weights are discarded.

\textbf{Phase 2} allocates the power budget across the mode's $k$ stages. For each slice, the cheapest
per-stage candidate reaching $\sigma \ge 1$ is chosen, unspent budget is
redistributed to the weakest stage, and the joint quality of the resulting
configuration is its bottleneck,
\begin{equation}
    \sigma_{\mathrm{joint}} = \min_{i=1,\dots,k} \sigma_i .
\end{equation}
Unique configurations are ranked by $\sigma_{\mathrm{joint}}$ and only the
top-$K$ are actually simulated end-to-end.

The winning configuration for each mode is the simulated candidate with the
highest achieved throughput.
\begin{equation}
    \mathrm{cfg}_{\mathcal{M}}^{\star} = \operatorname*{arg\,max}_{\mathrm{cfg}\,\in\,\mathrm{top}\text{-}K} \mathrm{tok/s}(\mathrm{cfg}),
\end{equation}

The two searches compose into one optimal-system-design pipeline: the
quantization search fixes $\mathbf{q}^\star$, the accuracy-validated
per-stage precision; the hardware/parallelism search then takes $\mathbf{q}^\star$ as
$\mathrm{dtype\_bytes}$ per stage and finds the throughput-optimal
hardware and parallelism configuration.

\section{Evaluation}
We evaluate PDAF disaggregation across a range of workload and hardware
characteristics to understand where and why its benefits hold. We first sweep
the disaggregation modes over the custom NPU design space
(\Cref{sec:npu-sweep}) and then repeat the sweep over commercially available
GPUs (\Cref{sec:gpu-sweep}), which lets us separate what disaggregation buys
from what stage-specialized hardware buys. \Cref{sec:allocation} inspects the
per-stage hardware the search selects in each case to explain the difference.
Finally, we isolate the contribution of individual model-architecture choices
through a controlled ablation (\Cref{sec:ablation-matrix}) and characterize the
accuracy cost of per-stage quantization (\Cref{sec:quant-sensitivity}).

\subsection{Experimental setup}

\textbf{Models.} The experiments span eight models across dense and
mixture-of-experts architectures at a range of scales, from
Llama-3.1-405B~\cite{grattafiori2024llama3herdmodels} and
DeepSeek-V4~\cite{deepseekv4} to GPT-OSS~\cite{gptoss2025}, Llama4-Scout,
Llama4-Maverick~\cite{llama4}, Qwen3-235B-A22B~\cite{qwen3}, and
GLM-4.6~\cite{glm46}.

\textbf{Workload.} The workload is parameterized by the prefill/output token
ratio ($I/O$) with output length held fixed at 1000 tokens and input length
scaled accordingly; per-request input and output lengths are then drawn from a
normal distribution centered on these targets to introduce realistic
variability. The length of the output sequences has been determined by running
BFCL~\cite{patil2025bfcl} and GSM8K~\cite{gsm8k} on real models. For each
config we simulate 500 requests at a fixed rate of 125 requests/second. The
setup was determined experimentally to provide high utilization of the hardware
without overloading.

\textbf{Disaggregation modes.} We compare four modes: \textbf{ND}
(non-disaggregated) co-locates all four stages on a single homogeneous pool;
\textbf{PD} disaggregates prefill from decode; \textbf{AF} disaggregates
attention from FFN; \textbf{PDAF} applies both splits, yielding four
independently provisioned stages (prefill-attention, prefill-FFN,
decode-attention, decode-FFN).

The sweeps of \Cref{sec:npu-sweep} and \Cref{sec:gpu-sweep} have been run with
full precision models; the compute cost of quantizing MoE models was too high
to run the search.

\subsection{Comparison of disaggregation modes for NPUs}
\label{sec:npu-sweep}

The NPU search explores a synthetic device design space
(\Cref{tab:design-space}): compute and memory technology are varied
independently and combined freely up to a fixed installed power budget. We
assume each disaggregated cluster is placed on NVSwitch~\cite{nvidia_nvlink} --
allowing up to 72 devices of 3600 GB/s bidirectional bandwidth -- and that the
NVSwitch instances are connected with Infiniband of 50 GB/s. Devices in
different disaggregation stages may differ: the Vera-Rubin platform (Groq LPU
for prefill and Rubin GPU for decode) falls within our search space.

\begin{figure*}[tbp]
    \centering
    \begin{subfigure}[t]{0.32\linewidth}
        \centering
        \includegraphics[width=\linewidth]{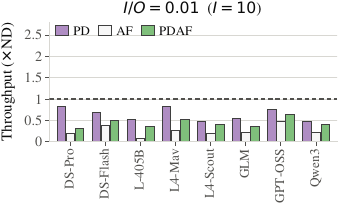}
        \caption{Decode-dominated.}
        \label{fig:npu_sweep_a}
    \end{subfigure}
    \hfill
    \begin{subfigure}[t]{0.32\linewidth}
        \centering
        \includegraphics[width=\linewidth]{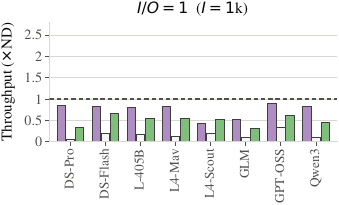}
        \caption{Balanced.}
        \label{fig:npu_sweep_b}
    \end{subfigure}
    \hfill
    \begin{subfigure}[t]{0.32\linewidth}
        \centering
        \includegraphics[width=\linewidth]{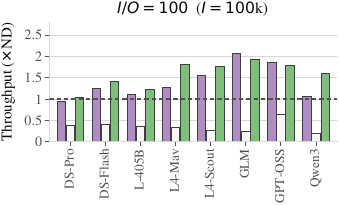}
        \caption{Prefill-dominated.}
        \label{fig:npu_sweep_c}
    \end{subfigure}
    \caption{NPU throughput relative to No Disaggregation ND ($\times$ND) for
    each disaggregation mode across models and workload profiles ($O{=}1000$
    tokens fixed; $I/O$ varies the prefill length). The dashed line marks ND
    parity, which is $1.00\times$ by definition; bars above it beat
    non-disaggregated serving. All three panels share a common $y$-axis.
    Models, left to right: DeepSeek-V4-Pro, DeepSeek-V4-Flash, Llama-3.1-405B,
    Llama-4-Maverick, Llama-4-Scout, GLM-4.6, GPT-OSS, Qwen3-235B-A22B.}
    \label{fig:npu_sweep}
    
\end{figure*}

\Cref{fig:npu_sweep} shows the ranking of serving strategies
at each workload ratio. At low and balanced prefill/output ratios ($I/O = 0.01$
and $I/O = 1$), ND is the strongest configuration for every model tested: PD,
AF, and PDAF all remain below $1.00\times$ ND across the board, with PD the
closest (up to $0.89\times$ for GPT-OSS at $I/O = 1$) but never crossing
parity. Disaggregation overhead simply outweighs any benefit when prefill work
is small relative to decode.

At $I/O = 100$, the ranking shifts. Both disaggregated prefill/decode
strategies clear parity almost universally: PDAF exceeds ND for all eight
models ($1.05$--$1.92\times$) and PD for seven of eight, with DeepSeek-V4-Pro
($0.95\times$) the sole configuration still below ND. PDAF is the best mode for
six of the eight models, reaching $1.81\times$ and $1.77\times$ for Llama-4
Maverick and Scout. AF remains the weakest mode at every ratio tested, never
exceeding ND; at $I/O = 100$ it spans $0.20$--$0.65\times$, its best showing but
still far from parity.

\begin{figure}[tbp]
    \centering
    \includegraphics[width=\linewidth]{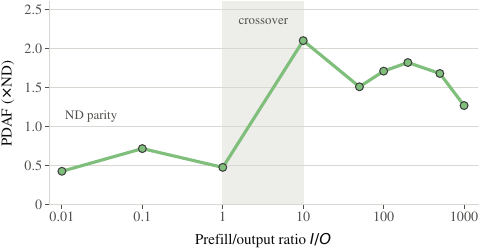}
    \caption{PDAF throughput relative to ND, averaged across all models in the
    NPU sweep, over nine prefill/output ratios ($O{=}1000$ tokens fixed). ND is
    $1.00\times$ by definition. The shaded decade brackets the crossover: PDAF
    is below parity at $I/O=1$ and well above it at $I/O=10$. Markers are drawn
    at all nine swept ratios; the axis labels decades only.}
    \label{fig:ratio_sweep}
\end{figure}

The three ratios of \Cref{fig:npu_sweep} bracket the crossover into a
disaggregation-favoring regime but do not locate it. \Cref{fig:ratio_sweep}
sweeps nine ratios and reports PDAF's benefit averaged across models. The
transition is abrupt and confined to a single decade: PDAF is still at
$0.48\times$ at $I/O=1$ and already at $2.10\times$ by $I/O=10$. It is also not
monotonic thereafter. The benefit plateaus between $1.5\times$ and $1.8\times$
through $I/O=500$ and then falls to $1.27\times$ at $I/O=1000$, consistent with
prefill growing large enough that the decode stages it feeds are no longer the
system bottleneck, so hardware provisioned separately for them is
increasingly idle. Agentic workloads sit near the middle of this range rather
than at its extremes, which is where the four-way split is most productive.

\begin{table*}[t]
\centering
\setlength{\tabcolsep}{2.5pt}
\renewcommand{\arraystretch}{1.15}
\footnotesize
\caption{%
  Per-stage device allocation for PDAF disaggregation at $I/O=100$~($I=100\mathrm{k}$, $O=1\mathrm{k}$),
  comparing the NPU design space against commercial GPU clusters.
  Each stage lists the device count $N$, the per-device memory operating point
  (technology, capacity in GB / bandwidth in GB\,s$^{-1}$), and peak compute in TFLOPS;
  the value beside each row label is the total device count for that cluster.
  The NPU search moves memory and compute independently -- decode stages take the top
  memory tier ($190$/$8.0$k) at the two extremes of the compute range, prefill stages take
  the top compute tier ($20$k) at the smallest capacities. The GPU catalogue ties the two
  together, so every stage receives the same operating point and only $N$ can vary.}
\label{tab:hw_comparison_pdaf_ratio100_merged}
\begin{tabular}{ll|rlr rlr rlr rlr}
\toprule
\textbf{Model} & & \multicolumn{3}{c}{\textbf{PA}} & \multicolumn{3}{c}{\textbf{PF}} & \multicolumn{3}{c}{\textbf{DA}} & \multicolumn{3}{c}{\textbf{DF}} \\
\cmidrule(lr){3-5}\cmidrule(lr){6-8}\cmidrule(lr){9-11}\cmidrule(lr){12-14}
 & & \textit{N} & \textit{Memory} & \textit{TFLOPS} & \textit{N} & \textit{Memory} & \textit{TFLOPS} & \textit{N} & \textit{Memory} & \textit{TFLOPS} & \textit{N} & \textit{Memory} & \textit{TFLOPS} \\
\midrule
\multirow{2}{*}{\textbf{DS-V4-Pro}}
 & \textit{NPU}\,(80)  & 16 & HBM 16/1.0k & 20k & 16 & HBM 140/4.8k & 20k & 24 & HBM 190/8.0k & 250 & 24 & HBM 190/8.0k & 2.5k \\
 & \textit{GPU}\,(100) & 32 & H100-80/3.4k & 989 & 32 & H100-80/3.4k & 989 & 4 & H100-80/3.4k & 989 & 32 & H100-80/3.4k & 989 \\
\addlinespace[2pt]\midrule
\multirow{2}{*}{\textbf{DS-V4-Flash}}
 & \textit{NPU}\,(80)  & 24 & HBM 16/1.0k & 20k & 8 & HBM 80/3.5k & 20k & 24 & HBM 190/8.0k & 250 & 24 & HBM 190/8.0k & 2.5k \\
 & \textit{GPU}\,(112) & 64 & H100-80/3.4k & 989 & 16 & H100-80/3.4k & 989 & 16 & H100-80/3.4k & 989 & 16 & H100-80/3.4k & 989 \\
\addlinespace[2pt]\midrule
\multirow{2}{*}{\textbf{L-405B}}
 & \textit{NPU}\,(88)  & 24 & LPDDR 32/200 & 20k & 8 & HBM 80/3.5k & 20k & 32 & HBM 190/8.0k & 100 & 24 & HBM 190/8.0k & 2.5k \\
 & \textit{GPU}\,(112) & 64 & H100-80/3.4k & 989 & 16 & H100-80/3.4k & 989 & 16 & H100-80/3.4k & 989 & 16 & H100-80/3.4k & 989 \\
\addlinespace[2pt]\midrule
\multirow{2}{*}{\textbf{L4-Mav}}
 & \textit{NPU}\,(108) & 16 & HBM 16/1.0k & 20k & 4 & HBM 140/4.8k & 20k & 64 & HBM 190/8.0k & 100 & 24 & HBM 190/8.0k & 2.5k \\
 & \textit{GPU}\,(112) & 32 & H100-80/3.4k & 989 & 16 & H100-80/3.4k & 989 & 32 & H100-80/3.4k & 989 & 32 & H100-80/3.4k & 989 \\
\addlinespace[2pt]\midrule
\multirow{2}{*}{\textbf{L4-Scout}}
 & \textit{NPU}\,(104) & 16 & HBM 16/1.0k & 20k & 8 & LPDDR 32/200 & 20k & 64 & HBM 190/8.0k & 100 & 16 & HBM 190/8.0k & 2.5k \\
 & \textit{GPU}\,(112) & 32 & H100-80/3.4k & 989 & 16 & H100-80/3.4k & 989 & 32 & H100-80/3.4k & 989 & 32 & H100-80/3.4k & 989 \\
\addlinespace[2pt]\midrule
\multirow{2}{*}{\textbf{GLM-4.6}}
 & \textit{NPU}\,(104) & 16 & HBM 16/1.0k & 20k & 8 & HBM 80/3.5k & 20k & 64 & HBM 190/8.0k & 100 & 16 & HBM 190/8.0k & 2.5k \\
 & \textit{GPU}\,(112) & 32 & H100-80/3.4k & 989 & 16 & H100-80/3.4k & 989 & 32 & H100-80/3.4k & 989 & 32 & H100-80/3.4k & 989 \\
\addlinespace[2pt]\midrule
\multirow{2}{*}{\textbf{GPT-OSS}}
 & \textit{NPU}\,(113) & 16 & HBM 16/1.0k & 10k & 1 & HBM 190/8.0k & 20k & 64 & HBM 190/8.0k & 25 & 32 & HBM 190/8.0k & 2.5k \\
 & \textit{GPU}\,(120) & 32 & H100-80/3.4k & 989 & 8 & A100-40/1.6k & 312 & 64 & H100-80/3.4k & 989 & 16 & H100-80/3.4k & 989 \\
\addlinespace[2pt]\midrule
\multirow{2}{*}{\textbf{Qwen3-235B}}
 & \textit{NPU}\,(92)  & 24 & HBM 40/2.0k & 20k & 4 & HBM 80/3.5k & 20k & 48 & HBM 190/8.0k & 100 & 16 & HBM 190/8.0k & 2.5k \\
 & \textit{GPU}\,(112) & 64 & H100-80/3.4k & 989 & 16 & H100-80/3.4k & 989 & 16 & H100-80/3.4k & 989 & 16 & H100-80/3.4k & 989 \\
\bottomrule
\end{tabular}
\end{table*}

\subsection{Comparison of disaggregations for heterogeneous GPUs}
\label{sec:gpu-sweep}

We repeat the sweep across commercially available hardware: each stage pool is
built from one or more identical AWS EC2 GPU instances (H100, A100, L40S, L4,
A10G, T4, V100, M60), with datasheet-accurate compute, memory, and TDP, and is
constrained by a fixed hourly \emph{cost} budget (USD/hr) rather than a power
budget. The interconnect bandwidth here is not uniform: single-instance pools
use intra-node bandwidth (NVLink/PCIe), while multi-instance pools are
bottlenecked by the slower inter-node fabric (EFA/NIC). As in the NPU search,
each disaggregation stage may be assigned a different device -- in PDAF, every
stage (prefill-attention, prefill-FFN, decode-attention and decode-FFN) can use
a different GPU type.

\begin{figure*}[tbp]
    \centering
    \begin{subfigure}[t]{0.32\linewidth}
        \centering
        \includegraphics[width=\linewidth]{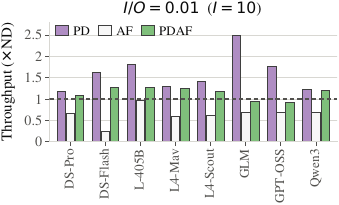}
        \caption{Decode-dominated.}
        \label{fig:gpu_sweep_a}
    \end{subfigure}
    \hfill
    \begin{subfigure}[t]{0.32\linewidth}
        \centering
        \includegraphics[width=\linewidth]{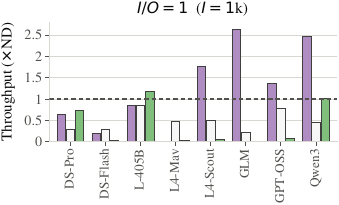}
        \caption{Balanced.}
        \label{fig:gpu_sweep_b}
    \end{subfigure}
    \hfill
    \begin{subfigure}[t]{0.32\linewidth}
        \centering
        \includegraphics[width=\linewidth]{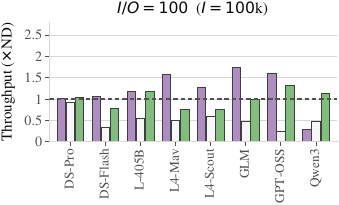}
        \caption{Prefill-dominated.}
        \label{fig:gpu_sweep_c}
    \end{subfigure}
    \caption{GPU throughput relative to No Disaggregation ND ($\times$ND) for
    each disaggregation mode across models and workload profiles ($O{=}1000$
    tokens fixed; $I/O$ varies the prefill length). Axes match
    \Cref{fig:npu_sweep} so the two design spaces can be compared directly.
    Models are ordered as in \Cref{fig:npu_sweep}.}
    \label{fig:gpu_sweep}
\end{figure*}

\Cref{fig:gpu_sweep} describes results for the GPU
experiment. The crossover into a disaggregation-favoring regime happens
earlier than on the NPU design space: already at $I/O=0.01$, PD exceeds ND for
all eight models ($1.18$--$2.50\times$) and wins outright in every case, where
the equivalent NPU sweep places ND ahead of every disaggregated mode at the
same ratio. The picture is also markedly less uniform. PD is above parity for
$8/8$ models at $I/O=0.01$ and $7/8$ at $I/O=100$, but only $4/8$ at $I/O=1$,
where it ranges from $0.01\times$ to $2.64\times$; PDAF is above parity for
$6/8$, $2/8$, and $4/8$ models at the three ratios respectively. AF never
exceeds ND at any ratio or for any model (best case $0.96\times$). Notably, the
finer four-way split is \emph{not} rewarded here: PD matches or outperforms
PDAF for six of eight models at $I/O=100$, the reverse of the NPU ranking at
the same ratio.

\subsection{Per-stage hardware allocation}
\label{sec:allocation}

\Cref{tab:hw_comparison_pdaf_ratio100_merged} shows the hardware
configuration the search selects for each stage at a prefill/output ratio of
100, and explains the divergence between the two design spaces.

Under traditional NPU technology, every stage uses HBM memory with only two
exceptions: Llama-3.1-405B's PA stage and Llama4-Scout's PF stage, which
instead use LPDDR. The specific HBM operating point still varies considerably
from stage to stage. The decode stages (DA, DF) always claim the highest
available capacity and bandwidth tier, but at opposite ends of the compute
range -- decode-attention is given the lowest compute tiers ($25$--$250$
TFLOPS) while decode-FFN takes $2.5$k; the prefill stages (PA, PF), by
contrast, primarily maximize compute at $20$k TFLOPS while accepting the
smallest capacity tiers.

The GPU allocations show no such spread, because GPUs fix compute and memory
bandwidth together rather than allowing them to be tuned independently per
stage. The search has little room to re-balance: $31$ of the $32$ PDAF stage
assignments are H100, the remaining one an A100, so nearly every stage receives
the same compute-to-bandwidth ratio regardless of whether it is attention- or
FFN-bound. With this reduced design-space flexibility, no single mode can
consistently match each stage's hardware to its own bottleneck, and the
relative benefit of PD, AF, and PDAF instead varies with the specific
compute/memory demands of each model and workload ratio. It also explains why
the four-way split does not pay off on GPUs: splitting attention from FFN only
helps when the two can be given genuinely different hardware, which is
precisely what the NPU design space permits and the GPU catalogue does not.

\subsection{Model architecture ablation}
\label{sec:ablation-matrix}

\begin{table}[t]
\centering
\small
\setlength{\tabcolsep}{5pt}
\caption{%
  Architecture ablation: PDAF throughput relative to ND ($\times$ND) at
  $I/O=100$, sweeping five factors on three MoE baselines spanning low, mid,
  and high decode-attention arithmetic intensity. Each factor varies one
  parameter with all others held at the baseline value: Precision sweeps
  \texttt{dtype\_bytes} (bytes per element), DA\_AI sweeps
  \texttt{kv\_lora\_rank} (latent width), Sparsity sweeps
  \texttt{num\_active\_experts}, Capacity sweeps \texttt{num\_experts}, and
  FFN size sweeps \texttt{ffn\_expansion}, the latter three as multiples of
  each model's baseline.
  \textbf{Bold} marks the best swept value per model per factor.}
\label{tab:ablation_matrix}
\begin{tabular}{llrrrr}
\toprule
\textbf{Factor} & \textbf{Model} & \multicolumn{4}{c}{\textbf{Swept value} $\rightarrow$ PDAF} \\
\midrule
 &  & 0.5 & 1 & 2 & 4 \\
\cmidrule(l){3-6}
\multirow{3}{*}{Precision} & GPT-OSS & \textbf{2.43} & 1.63 & 1.80 & 1.54 \\
 & GLM-4.6 & 1.29 & 1.29 & 1.92 & \textbf{2.20} \\
 & DS-V4-Flash & 0.85 & 0.95 & 1.41 & \textbf{1.47} \\
\midrule
 &  & 128 & 512 & 1024 & 4096 \\
\cmidrule(l){3-6}
\multirow{3}{*}{DA\_AI} & GPT-OSS & 0.93 & 1.23 & 1.64 & \textbf{1.85} \\
 & GLM-4.6 & 0.81 & 0.87 & 1.50 & \textbf{2.00} \\
 & DS-V4-Flash & 1.17 & 1.41 & 1.80 & \textbf{1.96} \\
\midrule
 &  & 0.25$\times$ & 1$\times$ & 4$\times$ & 8$\times$ \\
\cmidrule(l){3-6}
\multirow{3}{*}{Sparsity} & GPT-OSS & 1.77 & 1.80 & \textbf{1.88} & 1.69 \\
 & GLM-4.6 & 1.80 & \textbf{1.92} & 1.60 & 0.78 \\
 & DS-V4-Flash & \textbf{1.98} & 1.41 & 0.43 & 0.38 \\
\midrule
 &  & 0.5$\times$ & 1$\times$ & 2$\times$ & 4$\times$ \\
\cmidrule(l){3-6}
\multirow{3}{*}{Capacity} & GPT-OSS & 1.07 & \textbf{1.80} & 1.70 & 1.75 \\
 & GLM-4.6 & 1.92 & 1.92 & 1.92 & \textbf{1.98} \\
 & DS-V4-Flash & \textbf{1.42} & 1.41 & 1.41 & 1.41 \\
\midrule
 &  & 0.25$\times$ & 0.5$\times$ & 1$\times$ & 2$\times$ \\
\cmidrule(l){3-6}
\multirow{3}{*}{FFN size} & GPT-OSS & 1.75 & \textbf{2.00} & 1.80 & 1.57 \\
 & GLM-4.6 & \textbf{2.00} & 1.87 & 1.92 & 1.63 \\
 & DS-V4-Flash & \textbf{1.92} & 1.64 & 1.41 & 1.33 \\
\bottomrule
\end{tabular}
\end{table}

To isolate individual mechanisms, we ran a controlled ablation matrix: five
architecture factors, each swept independently while holding every other model
parameter fixed, repeated across three baseline models. Each factor is swept
over four points, and every configuration is evaluated at a fixed
prefill/output ratio of 100 under the same fixed hardware search and
installed-power budget used elsewhere in this work. The factors are the
following:

\begin{itemize}[leftmargin=1em]
  \item \textbf{Precision} - FP4 through FP32. The only lever that shifts every stage's
        arithmetic intensity at once.
  \item \textbf{Decode attention arithmetic intensity}  -
        MLA latent width, which sets KV bytes per token and hence
        decode-attention arithmetic intensity.
  \item \textbf{Sparsity} - experts activated per token, i.e.\ active
        FFN compute and the power contention it creates.
  \item \textbf{Capacity} - total FFN weight
        footprint with active compute pinned. The deliberate counterpart to
        Sparsity: it separates weight \emph{capacity} pressure from active
        \emph{compute} pressure, testing whether capacity growth alone is the
        cheaper of the two.
  \item \textbf{FFN size} -
        moves both the active work (as Sparsity does) and the resident
        footprint (as Capacity does). Since only the product
        $\mathrm{ex}\!\cdot\!k$ enters the active terms, doubling
        \texttt{ffn\_expansion} should match doubling
        \texttt{num\_active\_experts} unless the added footprint changes the
        outcome.
\end{itemize}

We report four findings.

{\textbf{Finding 1:} PDAF's benefit grows monotonically with decode-attention
KV traffic.} \texttt{kv\_lora\_rank} is the width of the compressed latent KV vector that MLA-style attention caches per token, and thus sets KV bytes per token per layer. Sweeping \texttt{kv\_lora\_rank}  from $128$ to $4096$ moves PDAF's
benefit monotonically upward in all three models, by $+100\%$ for GPT-OSS
($0.93\times\!\to\!1.85\times$), $+147\%$ for GLM-4.6
($0.81\times\!\to\!2.00\times$), and $+67\%$ for DeepSeekV4-Flash
($1.17\times\!\to\!1.96\times$). Note that this direction is one of
\emph{decreasing} decode-attention arithmetic intensity: FLOPs per decode step
are independent of the rank while KV bytes scale with it, so over this sweep
$\mathrm{DA\_AI}$ falls from $10.7$ to $1.8$. The mechanism is visible directly
in the roofline structure of decode-attention. Its per-step byte count carries a
KV term proportional to context length and a weight term that is independent of
batch, so its arithmetic intensity is set by the ratio between them. At
$\texttt{kv\_lora\_rank}=128$ the KV term is small enough that decode-attention
is weight-read dominated and therefore structurally indistinguishable from
decode-FFN: the two stages want the same hardware, the attention/FFN split has
nothing to separate, and PDAF falls \emph{below} the non-disaggregated baseline
for two of the three models ($0.93\times$, $0.81\times$). As the rank grows, KV
traffic comes to dominate and decode-attention's requirement diverges from
decode-FFN's, which is precisely the asymmetry the four-way split exists to
exploit.

{\textbf{Finding 2:} memory-capacity growth is essentially free.} Growing
\texttt{num\_experts} while holding \texttt{num\_active\_experts} fixed leaves
every stage's per-token compute and bandwidth demand unchanged; FLOPs and
weight-bytes-read in the decode-FFN roofline depend only on the number of
\emph{active} experts, and increasing the total raises only the capacity needed
to hold all experts' weights resident. PDAF's benefit is correspondingly
insensitive to it. Quadrupling the expert count relative to baseline moves
GLM-4.6 by $+3\%$ and DeepSeekV4-Flash by under $0.5\%$ (flat at $1.41\times$
at every point), indicating capacity never becomes the binding constraint for
these configurations. GPT-OSS is non-monotonic, and notably its \emph{worst}
point is the smallest expert count, $41\%$ below its baseline
($1.07\times$ versus $1.80\times$ at $0.5\times$ experts), while $2\times$ and
$4\times$ sit within $6\%$ of baseline.

{\textbf{Finding 3:} active-compute growth is the only factor that reduces
PDAF's benefit.} Growing \texttt{num\_active\_experts} raises decode-FFN's
per-token compute \emph{and} its per-step weight traffic together, and this is
the one lever that collapses PDAF's advantage. Relative to baseline,
DeepSeekV4-Flash loses $69\%$ at $4\times$ active experts
($1.41\times\!\to\!0.43\times$) and $73\%$ at $8\times$ ($0.38\times$). GLM-4.6
loses $17\%$ at $4\times$ ($1.60\times$) before collapsing by $60\%$ at
$8\times$ ($0.78\times$). Only GPT-OSS is insensitive, staying within $6\%$ of
baseline across the whole range ($1.69\times$--$1.88\times$); it is also the
model with the lowest baseline active-expert count, so the same relative
multiple leaves it at a lower absolute compute demand. Sweeping
\texttt{ffn\_expansion}, the other lever on the same
$\mathrm{ex}\!\cdot\!k$ compute proxy, reproduces the ordering but not the
magnitude: over $0.25\times$ to $2\times$, DeepSeekV4-Flash declines steadily
by $31\%$ ($1.92\times\!\to\!1.33\times$) while GLM-4.6 and GPT-OSS decline by
$19\%$ and $10\%$ --- an order of magnitude milder than the $73\%$ collapse the
active-expert lever produces.

{\textbf{Finding 4:} reduced precision is not uniformly beneficial.} Sweeping
\texttt{dtype\_bytes} splits the three models rather than moving them together.
Narrowing precision from $4$ to $0.5$ bytes per element improves GPT-OSS by
$+57\%$ ($1.54\times\!\to\!2.43\times$) but costs GLM-4.6 $41\%$
($2.20\times\!\to\!1.29\times$) and DeepSeekV4-Flash $42\%$
($1.47\times\!\to\!0.85\times$), the latter falling below parity with the
non-disaggregated baseline. Precision scales the weight-byte term of every stage
simultaneously, so narrowing it raises the arithmetic intensity of all four
stages at once; whether that helps PDAF depends on whether it widens or narrows
the \emph{gap} between decode-attention and decode-FFN, which is what the split
monetizes. For models whose decode-attention is already KV-dominated, shrinking
weight bytes leaves the KV term untouched and preserves the asymmetry; where
weight traffic is what distinguished the stages, removing it collapses the
distinction.

\subsection{Quantization sensitivity}
\label{sec:quant-sensitivity}

Evaluating the quantization search is considerably more expensive than the
throughput sweeps above, since each candidate assignment requires a full
accuracy evaluation rather than a single simulator pass. We therefore restrict
this analysis to a targeted sensitivity study on one smaller model,
Qwen3.5-32B, and characterize the accuracy cost of reducing precision at each
stage in isolation. We consider two prefill-dominated reasoning workloads,
BFCL and a subset of GSM8K, which differ substantially in shape: BFCL averages
$11{,}174$ prefill and $1{,}472$ decode tokens per request, against $995$ and
$313$ for GSM8K, an approximately $11\times$ difference in context length.

Beginning from an $8$-bit baseline applied uniformly to all four stages
($75\%$ on GSM8K, $21\%$ on BFCL), we reduce exactly one stage to $4$ bits and
re-evaluate. Uniform $4$-bit quantization collapses accuracy on both tasks
($11\%$ and $6\%$), confirming that $4$ bits is too aggressive when applied
globally. The single-stage reductions, however, expose a pronounced asymmetry
between stage classes. Reducing either FFN stage, prefill or decode, is
severely damaging on GSM8K ($47\%$ and $15\%$ accuracy respectively, the latter
comparable to the uniform $4$-bit collapse) while leaving BFCL essentially
unchanged ($20\%$ in both cases, at baseline). Reducing either attention stage
inverts the pattern: BFCL accuracy is approximately halved ($11$--$12\%$),
whereas GSM8K remains at or marginally above the $8$-bit baseline
($79$--$80\%$). Which stage tolerates low precision is therefore a property of
the workload rather than of the model alone, and a single global precision
choice necessarily sacrifices one of the two.

\begin{table}[th!]
\centering
\small
\caption{QWEN 32B quantization results on BFCL and GSM8k (20\%).}
\label{tab:qwen32b-quant}
\begin{tabular}{lcc}
\toprule
Configuration & BFCL & GSM8k (20\%) \\
\midrule
Baseline & 18\% & 78\% \\
8/8/8/8     & 21\% & 75\% \\
4/4/4/4     & 6\%  & 11\% \\
\midrule
8/4/8/8     & \textbf{20\%} & 47\% \\
8/8/4/8     & 11\% & \textbf{79\%} \\
4/8/8/8     & 12\% & \textbf{80\%} \\
8/8/8/4     & \textbf{20\%} & 15\% \\
\bottomrule
\end{tabular}
\end{table}

\section{Conclusions}
We present an event-driven simulator for stage-disaggregated LLM inference that
treats prefill-attention, prefill-FFN, decode-attention and decode-FFN as
independently provisionable stages, each with its own hardware, parallelism and
numeric precision, and searches the resulting design space under a power or
cost budget. We validate its component models against an $8\times$B200 node.

Using the simulator, we evaluate stage disaggregation on both commercial GPUs
and custom accelerators, and generalize the result across eight models and five
orders of magnitude of workload profile. A factorial ablation isolates which
architectural properties drive the benefit, and a quantization study shows that
the precision-sensitive stage is a property of the workload.

Several limitations bound these results. Validation covers the simulator's
components rather than its end-to-end serving behavior, leaving scheduling and
batching effects unverified, and the quantization study covers one model and two
tasks without repeated runs, so its stage asymmetry should be read as a
direction rather than a calibrated magnitude. The results of the GPU experiment
need to be considered with the fact that they are based on a single particular
inference provider offer and can vary with different pricing models.

These bounds point directly at the next steps. Validating the simulator against
a deployed disaggregated serving stack would close the gap between component
accuracy and end-to-end behavior, and is the prerequisite for trusting the
scheduling and batching effects the current model abstracts away. The
quantization search is the other open front: the accuracy evaluation, rather
than the simulator, is what makes it expensive, so extending it beyond one
model and four stage-wise assignments depends on cheaper accuracy proxies
rather than on faster simulation. Finally, the ablation suggests that PDAF's
benefit is largely predicted by a model's decode-attention arithmetic intensity
and its active FFN compute. If that relationship holds across a wider set of
architectures, the four-way split could be ruled in or out from a model's
configuration alone, without a hardware search at all.


\clearpage
\bibliographystyle{IEEEtranS}
\bibliography{ref}

\end{document}